\documentclass[final,1p,times,authoryear]{elsarticle}
\usepackage{amsmath,longtable,tikz,url,natbib}

\usetikzlibrary{arrows.meta}
\tikzset{>={latex[scale=2]}}
\usetikzlibrary{positioning}

\def\<#1>{\langle#1\rangle}

\let\set\mathbb
\def\i{\mathrm{i}}

\begin{document}

\begin{frontmatter}

  \title{Exploring the Meta Flip Graph for Matrix Multiplication\footnote{An earlier version of this work was presented in the poster ``Consequences of the Moosbauer-Poole Algorithms'' at ISSAC'25~\citep{kauers25}.
    M. Kauers was supported by the Austrian FWF grants 10.55776/PAT8258123, 10.55776/I6130, and 10.55776/PAT9952223.
    I. Wood was supported by the Austrian FWF grants 10.55776/PAT8258123.
    The article was finalized while the first author was attending the special semester on
    Complexity and Linear Algebra at the Simons Institute in Berkeley.     
    }
  }

\author{Manuel Kauers}
\address{Institute for Algebra $\cdot$ Johannes Kepler University, Linz, Austria}
\ead{manuel.kauers@jku.at}

\author{Isaac Wood}
\address{Institute for Algebra $\cdot$ Johannes Kepler University, Linz, Austria}
\ead{isaac.wood@jku.at}

\begin{abstract}
  Continuing recent investigations of bounding the tensor rank of matrix multiplication using
  flip graphs, we present here improved rank bounds for about thirty matrix formats. 
\end{abstract}


\end{frontmatter}

\section{Introduction}

During the past few years, we have seen renewed activity in the search for matrix
multiplication schemes of low rank for small and specific matrix multiplication
formats.  After Strassen's discovery that two $2\times 2$ matrices over any
coefficient ring can be multiplied using only seven multiplications, one less
than the standard algorithm, people have already been searching intensively for
multiplication schemes that multiply an $n\times m$ matrix with an $m\times p$
matrix over an arbitrary coefficient ring using as few multiplications as
possible. Notable early milestones in this effort are Laderman's
algorithm~\citep{La:Anaf}, which for $(n,m,k)=(3,3,3)$ uses 23 multiplications rather
than~27, and the paper by \cite{HK:OMtN}, which discusses
the ranks of all formats of the form $(2,m,p)$. 

The minimal number of multiplications required for multiplying an $n\times m$ matrix
with an $m\times p$ matrix is known as the \emph{rank} of the matrix multiplication
tensor of format $(n,m,p)$. By slight abuse of language, we also say that the number
of multiplications needed by a specific algorithm for carrying out this product is
the rank of this algorithm. Note that the rank of an algorithm is an upper bound for
the rank of the tensor. While most of the work on matrix multiplication focuses on
the tensor rank of asymtptotically large matrices, also the ranks for many small
matrix formats remain unknown. 

Nowadays computers are used for finding for matrix
multiplication schemes of low rank. There are several approaches. \cite{Sm:Tbca}
found several improvements using numerical
optimization. \cite{DBLP:journals/tcs/DrevetIS11} used a variant of
Pan's trilinear aggregation method~\citep{pan84} to find 
improvements for various medium-sized formats. \cite{sedoglavic17} uses a
divide-and-conquer technique to also improve various medium-size
formats. \cite{HKS:Nwtm,heule19a} used SAT solvers to find many new instances
matching Laderman's bound for $(3,3,3)$, but found no improvements. AlphaTensor
and AlphaEvolve~\citep{FBH+:Dfmm,novikov25} used machine learning to improve the
rank bounds for a number of small formats. \cite{kauers23f,kauers24c} introduced
flip graphs and found better ranks for $(5,5,5)$ by performing random searches
in this graph.

Here we continue the development of the flip graph approach.
As introduced by~\cite{kauers23f}, the flip graph for a certain fixed format $(n,m,k)$
is a graph whose vertices are all the multiplication algorithms for multiplying
an $n\times m$ matrix with an $m\times p$ matrix, regardless of how many multiplications
are needed. Two such algorithms are connected by an edge if it is possible to
reach one of them from the other by means of a certain transformation. There
are three kinds of transformations. The first one, called ``flip'', connects
two algorithms of the same rank. 
The second one, called a ``reduction'', leads from one algorithm to an algorithm
of rank one less.
The third transformation, which is called ``plus'' and was introduced by~\cite{arai24},
leads from one algorithm to an algorithm of rank one higher. 

In search for algorithms of low rank, reduction edges are most useful, but they are
rare. The flip graph method proceeds by starting from some given algorithm and traversing
a random path of flip edges until a node is encountered which has a reduction edge.
Then this edge is taken and the search continues. Following a plus edge is counter
productive because it increases the rank, but it is still useful to occasionally
choose such an edge in order to avoid getting trapped in a part of the graph from
which no further reduction edges can be reached.

The technical details of the flip graph search procedure are not important for the
present work. What matters is that for every format $(n,m,p)$, there is one flip
graph, and given a node, i.e., an algorithm for multiplying an $n\times m$ matrix
with an $m\times p$ matrix, there is a way to search for algorithms for the same
format but of smaller rank. In principle, an algorithm of minimal rank can always
be found in this way \cite[Thm. 9]{kauers23f}, but in practice, the graph is so large that
we can easily fail to find an optimal algorithm, and even if we do find one, it
cannot easily be recognized that there is no better one.

Whether or not a search succeeds depends partly on luck, partly on the choice of
certain parameters (e.g., how often to take a plus edge), partly on the amount
of computing power invested, and to a greater extent on the choice of the
starting point of the search. Already in their first paper on the subject,
\cite{kauers23f} noticed that while they were not able to reach a better rank
than 97 for $(5,5,5)$ when starting from the standard algorithm, they easily
reached a rank scheme of rank 95 (over $\set Z_2$) when they took as starting
point a scheme of rank~96 that had been found by AlphaTensor~\citep{FBH+:Dfmm}.

One year later, \cite{arai24} found a scheme of rank 94 for $(5,5,5)$ (over
$\set Z_2$) using an incremental approach. Starting from the standard algorithm
for $(2,2,2)$, they apply a flip graph search to get a scheme of rank~7. Then
they extend this scheme to a scheme of rank 11 for $(2,2,3)$ by exploiting the
block multiplication rule $A(B|C)=(AB|AC)$ for matrices and use it as starting
point for another new flip graph search. They proceed in the same way to
construct good schemes of formats $(2,3,3)$, $(3,3,3)$, $(3,3,4)$, $(3,4,4)$,
$(4,4,4)$, $(4,4,5)$, $(4,5,5)$, and finally $(5,5,5)$.

The approach of \cite{arai24} not only led to improved bounds for $(4,5,5)$ and $(5,5,5)$,
but also required substantially less computation time than the searches which
use the standard algorithm as starting point.
It thus seems to be a good idea to construct starting points from optimized
schemes of neighboring formats.
For the present paper, we have explored this approach further.

We connect the flip graphs for the various formats $(n,m,p)$ to one single graph by
adding two new kinds of edges which connect schemes of different formats.
The first kind is called \emph{extension.} It connects every multiplication
scheme of some format $(n,m,p)$ to a scheme of format $(n,m,p+1)$ or
$(n,m+1,p)$ or $(n+1,m,p)$. This is the same operation that was used by
\cite{arai24}.

The second kind of new edges is called \emph{projection}. It connects every
multiplication scheme of some format $(n,m,p)$ to a scheme of format $(n,m,p-1)$
or $(n,m-1,p)$ or $(n-1,m,p)$ by simply setting all the variables corresponding to
a certain row or column to zero. For example, if we have a $(2,2,3)$ multiplication
scheme for computing the product
\[
\begin{pmatrix}
  a_{1,1} & a_{1,2} \\
  a_{2,1} & a_{2,2} 
\end{pmatrix}
\begin{pmatrix}
  b_{1,1} & b_{1,2} & b_{1,3} \\
  b_{2,1} & b_{2,2} & b_{2,3}
\end{pmatrix} =
\begin{pmatrix}
  c_{1,1} & c_{1,2} & c_{1,3} \\
  c_{2,1} & c_{2,2} & c_{2,3}
\end{pmatrix},
\]
then we can obtain a $(2,2,2)$ multiplication scheme by setting $b_{1,3}$ and $b_{2,3}$
to zero and dropping the variables $c_{1,3}$ and $c_{2,3}$.

The graph obtained from all the flip graphs by adding all extension and projection edges
is called the \emph{meta flip graph.} The purpose of this article is to share a bunch
of rank improvements that were obtained by exploring this graph.

\section{Searching From Scratch}\label{sec:2}

Given a collection of multiplication schemes of some format $(n,m,p)$ and some rank~$r$,
here is what we do to try to reduce the rank.
We consider $100nmp$ random paths in the flip graph that start at randomly chosen starting points
chosen from the given collection.
The length of each path is restricted to $100000nmp$.
Most of the edges in the paths are flips, occasionally we also allow plus edges.
If a vertex that has a reduction edge, then we do the reduction.
As soon as we reach a vertex whose rank is $r-1$, i.e., as soon as the number of
reduction steps taken exceeds the number of plus steps taken, we abort the search
on this specific path and record the resulting scheme.

The result of this process is a collection of various schemes of rank $r-1$,
with which we proceed in the same way to obtain a collection of schemes of rank
$r-2$. The process is repeated until we reach a rank at which none of the
considered random paths leads to a reduction.

In order to traverse the meta flip graph, where we also allow extensions and
projections to change the format of the multiplication algorithms under
consideration, we proceed as follows.

We take the standard algorithm for format $(2,2,2)$ as starting point,
apply the procedure outlined above, and arrive at a collection of schemes
of rank~7. From all these schemes, we construct schemes of format $(2,2,3)$
by extension. Because of the symmetry of the matrix multiplication tensor,
there is no need to also consider the formats $(2,3,2)$ or $(3,2,2)$. Also,
there is no need to consider projections, because formats $(n,m,p)$ with
$\min(n,m,p)=1$ are trivial.

With the schemes of format $(2,2,3)$, we apply again the flip graph search
procedure to find schemes of smallest possible rank. The resulting collection
of schemes is then extended into a collection of schemes of format $(2,2,4)$
and a collection of schemes of format $(2,3,3)$. At this point, we could in
principle also already apply a projection to return to format $(2,2,2)$, but
we chose to refrain from returning to formats where we have already been.

Continuing in this way, we have explored most of the paths in the grid of
formats $(n,m,p)$ of length 11 that start at $(2,2,2)$ and don't visit the same
format more than once. There are altogether 1245 such paths, and each format
$(n,m,p)$ with $2\leq\min(n,m,p)$ and $\max(n,m,p)\leq8$ and $n+m+k\leq14$
except for $(n,m,p)=(4,4,6)$ is visited by at least one of them.

\begin{figure}
  \begin{center}
    \sffamily
    \begin{tikzpicture}[x={(0cm,1cm)},z={(.6cm,.4cm)},y={(1cm,0cm)},scale=1.6,rounded corners=2pt,inner sep=2pt,outer sep=1pt]
\node (233) at (2,3,3) {\footnotesize 233};
\node (247) at (2,4,7) {\footnotesize 247};
\node (255) at (2,5,5) {\footnotesize 255};
\node (225) at (2,2,5) {\footnotesize 225};
\node (256) at (2,5,6) {\footnotesize 256};
\node (234) at (2,3,4) {\footnotesize 234};
\node (235) at (2,3,5) {\footnotesize 235};
\node[draw] (257) at (2,5,7) {\footnotesize 257};
\node (266) at (2,6,6) {\footnotesize 266};
\node (224) at (2,2,4) {\footnotesize 224};
\node (223) at (2,2,3) {\footnotesize 223};
\node[draw] (268) at (2,6,8) {\footnotesize 268};
\node (236) at (2,3,6) {\footnotesize 236};
\node (245) at (2,4,5) {\footnotesize 245};
\node[draw] (267) at (2,6,7) {\footnotesize 267};
\node (244) at (2,4,4) {\footnotesize 244};
\node[draw,fill=black] (222) at (2,2,2) {\footnotesize \textcolor{white}{222}};
\node (246) at (2,4,6) {\footnotesize 246};
\draw[->](234)--(244);
\draw[->](246)--(256);
\draw[->](223)--(233);
\draw[->](246)--(247);
\draw[->](245)--(255);
\draw[->](266)--(267);
\draw[->](255)--(256);
\draw[->](267)--(268);
\draw[->](247)--(257);
\draw[->](235)--(236);
\draw[->](256)--(266);
\draw[->](233)--(234);
\draw[->](235)--(245);
\draw[->](244)--(245);
\draw[->](245)--(246);
\draw[->](223)--(224);
\draw[->](225)--(235);
\draw[->](234)--(235);
\draw[->](224)--(225);
\draw[->](222)--(223);
\draw[->](236)--(246);
\draw[->](224)--(234);
\end{tikzpicture}
  \end{center}
  \caption{Paths in the meta flip graph starting from the standard algorithm for format $(2,2,2)$ (black)
    that led to improvements for certain formats (boxed).}\label{fig:bottomup}
\end{figure}
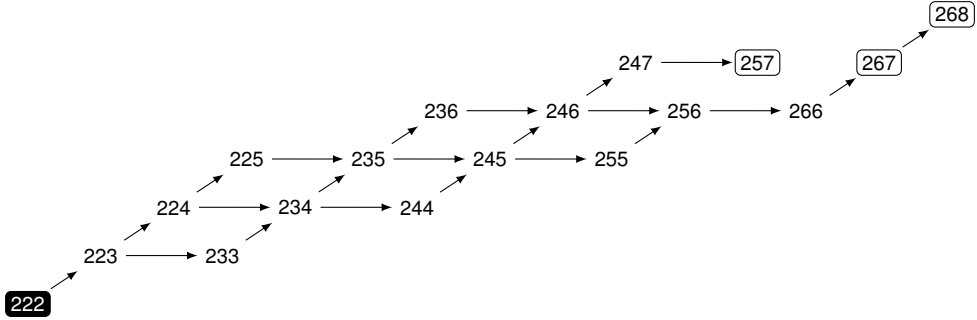

For the formats $(2,5,7)$, $(2,6,7)$, and $(2,6,8)$, we were able to improve the rank bound
by one.
In the case of $(2,6,7)$, there were seven paths that led to the improvement.
For each of the two other formats, only one path led to the improvement.
Figure~\ref{fig:bottomup} shows all the nodes visited by some path that led to
an improvement.
We also considered paths involving projection steps, but it turned out that
none of the paths that led to an improvement contained such an edge. 

\section{Using other Starting Points}\label{sec:3}

Although we do not know for sure, it is fair to believe that for most of the formats $(n,m,p)$ near $(2,2,2)$,
the currently best known upper bounds for the rank are actually sharp. This would explain why starting from
$(2,2,2)$, we have to go some distance before reaching a format where a rank drop can be achieved. However,
continuing this approach to even larger format becomes prohibitively expensive as the number of paths grows
quickly with their lengths.

An attractive alternative is to take starting points of formats other than $(2,2,2)$.
Multiplication schemes that were obtained by other techniques are natural choices.
For example, \cite{moosbauer25} have recently obtained substantial rank improvements for the formats
$(5,5,5)$ and $(6,6,6)$ by using a variant of the flip graph method that takes symmetries
of the matrix multiplication tensor into account.
Applying a meta flip graph search as described in the previous section with their schemes
as starting points, we were able to obtain rank reductions for a number of formats in
the vicinity of $(5,5,5)$ and $(6,6,6)$, respectively.

\begin{figure}
  \begin{center}
    \sffamily    
    \begin{tikzpicture}[x={(1cm,0cm)},z={(.4cm,.7cm)},y={(0cm,1cm)},scale=1.8,rounded corners=2pt,inner sep=2pt,outer sep=3pt]
\node (456) at (4,5,6) {\footnotesize 456};
\node[draw] (577) at (5,7,7) {\footnotesize 577};
\node[draw] (556) at (5,5,6) {\footnotesize 556};
\node (445) at (4,4,5) {\footnotesize 445};
\node (455) at (4,5,5) {\footnotesize 455};
\node[draw] (446) at (4,4,6) {\footnotesize 446};
\node[draw] (567) at (5,6,7) {\footnotesize 567};
\node[draw] (566) at (5,6,6) {\footnotesize 566};
\node[draw] (457) at (4,5,7) {\footnotesize 457};
\node (466) at (4,6,6) {\footnotesize 466};
\node[draw] (568) at (5,6,8) {\footnotesize 568};
\node[draw] (458) at (4,5,8) {\footnotesize 458};
\node[draw] (467) at (4,6,7) {\footnotesize 467};
\node[draw,fill=black] (555) at (5,5,5) {\footnotesize \textcolor{white}{555}};
\node[draw] (558) at (5,5,8) {\footnotesize 558};
\node[draw] (557) at (5,5,7) {\footnotesize 557};
\draw[->](456)--(556);
\draw[->](557)--(558);
\draw[->](457)--(458);
\draw[->](457)--(467);
\draw[->](558)--(458);
\draw[->](567)--(557);
\draw[->](566)--(466);
\draw[->](567)--(568);
\draw[->](555)--(556);
\draw[->](566)--(567);
\draw[->](556)--(456);
\draw[->](455)--(445);
\draw[->](567)--(577);
\draw[->](556)--(566);
\draw[->](455)--(456);
\draw[->](456)--(466);
\draw[->](445)--(446);
\draw[->](557)--(567);
\draw[->](456)--(446);
\draw[->](557)--(457);
\draw[->](456)--(457);
\draw[->](555)--(455);
\draw[->](466)--(467);
\draw[->](556)--(557);
    \end{tikzpicture}\hfil
%
% moosbauerpoole bB
\begin{tikzpicture}[x={(1cm,0cm)},z={(.4cm,.7cm)},y={(0cm,1cm)},scale=1.8,rounded corners=2pt,inner sep=2pt,outer sep=3pt]
\node (667) at (6,6,7) {\footnotesize 667};
\node[draw,fill=black] (666) at (6,6,6) {\footnotesize \textcolor{white}{666}};
\node[draw] (567) at (5,6,7) {\footnotesize 567};
\node[draw] (556) at (5,5,6) {\footnotesize 556};
\node[draw] (557) at (5,5,7) {\footnotesize 557};
\node[draw] (577) at (5,7,7) {\footnotesize 577};
\node[draw] (566) at (5,6,6) {\footnotesize 566};
\node (466) at (4,6,6) {\footnotesize 466};
\node[draw] (568) at (5,6,8) {\footnotesize 568};
\node[draw] (467) at (4,6,7) {\footnotesize 467};
\node (668) at (6,6,8) {\footnotesize 668};
\draw[->](666)--(566);
\draw[->](667)--(668);
\draw[->](566)--(556);
\draw[->](567)--(577);
\draw[->](566)--(466);
\draw[->](567)--(566);
\draw[->](466)--(467);
\draw[->](567)--(467);
\draw[->](567)--(557);
\draw[->](567)--(568);
\draw[->](668)--(568);
\draw[->](566)--(567);
\draw[->](667)--(567);
\draw[->](556)--(557);
\draw[->](666)--(667);
\end{tikzpicture}
\end{center}
\caption{Paths in the meta flip graph starting from a scheme found by~\cite{moosbauer25} (black)
  that led to improvements for certain formats (boxed).}\label{fig:moosbauerpoole}
\end{figure}
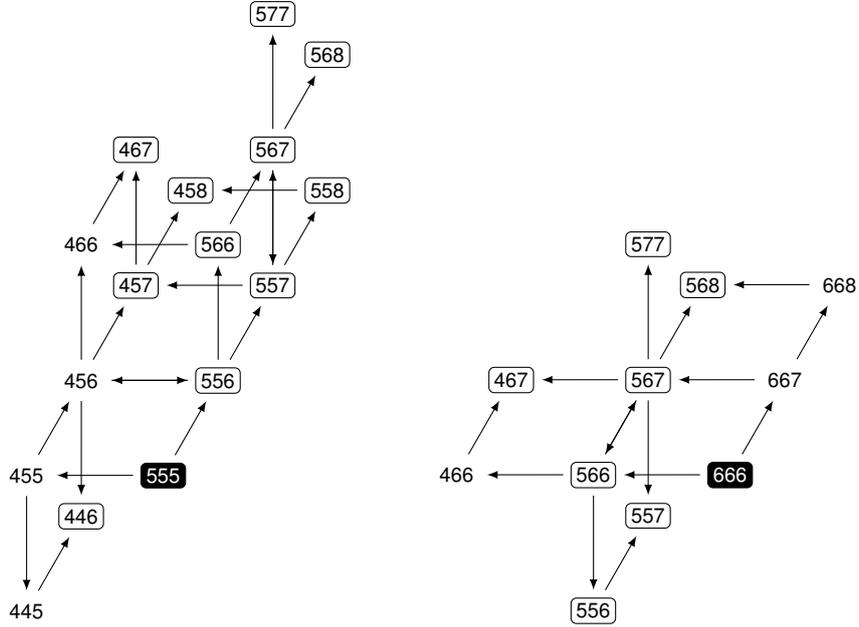

Figure~\ref{fig:moosbauerpoole} shows the formats visited on paths that led to success.
These improvements were already announced on our ISSAC'25 poster~\citep{kauers25}.
Note that unlike in the previous section, there are now also successful paths containing
projection edges.
Another difference to the approach from Section~\ref{sec:2} is that some improvements
are obtained via paths that pass through suboptimal vertices.
For example, starting from $(5,5,5)$, there are three paths reaching rank 123
for $(4,6,7)$ via the vertex $(4,6,6)$, although the search only found schemes
of rank 106 for this format. The best known rank bound for $(4,6,6)$ is 105.
Altogether, eight of the 25 paths leading to a rank improvement pass through a
vertex for which the best known rank bound was not reached. 

Shortly after the preprint of our ISSAC poster abstract appeared on ArXiv,
AlphaEvolve \citep{novikov25} released rank improvements for about a dozen
formats. Taking these as starting points and exploring their vicinity in the
meta flip graph, we were able to identify some further rank improvements.

\section{Coefficient Domains with more than Two Elements}\label{sec:4}

Turning from one algorithm to another algorithm via a flip amounts to adding
something and subtracting it at some other position. In order to avoid having to
decide what to add and subtract, flip graph searches have so far always been
performed with $\set Z_2$ as ground domain. This has the advantage that there is
only one choice what to add and subtract (because adding and subtracting zero
would not change anything). The disadvantage of taking $\set Z_2$ as a ground
domain is that at the end of the search, we only obtain a multiplication
algorithm that is valid for this domain. It would be more interesting to have
algorithms with coefficients in $\set Z$ or at least in~$\set Q$, which are
applicable to matrices over any coefficient domain in which the denominators of
the fractions appearing in the coefficients are invertible.

Hensel lifting~\citep{vG:MCA} can be used in order to translate an algorithm for the
ground domain $\set Z_2$ into an algorithm for the ground domain $\set
Z_{2^\ell}$, for some $\ell\in\set N$. Every increment of the exponent $\ell$
requires solving an inhomogeneous linear system over $\set Z_2$. Occasionally,
this system happens to be unsolvable, and in this case, no extension to a higher
exponent is possible. In most of the cases however, we observe that the
algorithms found by the flip graph method with the ground domain $\set Z_2$ can
be lifted to $\set Z_{2^\ell}$ for a sufficiently large $\ell$ such that
rational reconstruction \citep{vG:MCA} succeeds turning the algorithm into one for
the ground domain~$\set Q$.

Algorithms with coefficients in $\set Q$ that involve one or more fractions with
even denominator cannot be found in this way, because such fractions have no
valid homomorphic image in~$\set Z_2$. For this reason, in the computations
reported in the previous section we could only explore the neighborhoods of 9 of
the 13 new algorithms found by AlphaEvolve. The other four had coefficients in
$\set Z[\frac12]$ or in $\set Z[\frac12,\i]$. \cite{dumas25,dumas25a} succeeded
in eliminating $\i$ from these schemes, but not~$\frac12$. 

In order to also explore the neighborhoods of the schemes that are not valid
mod~2, we implemented the flip graph method for the ground domains $\set Z_3$
and $\set Z_5$. They both contain multiplicative inverses of~$2$, and the latter
in addition contains a root of the polynomial $x^2+1$.  Unfortunately, these
efforts were in vein. While in some cases we succeeded in matching the currently
best known rank bound, we did not encounter any improvement.

\section{Another Search from Scratch}\label{sec:5}

In an extension step, one of the three dimensions of a matrix multiplication format $(n,m,p)$
is increased by one. For example, an algorithm for format $(3,4,6)$ can be obtained from
an algorithm for format $(3,4,5)$ by observing that the matrix product
\[
\begin{pmatrix}
  a_{1,1} & a_{1,2} & a_{1,3} & a_{1,4} \\
  a_{2,1} & a_{2,2} & a_{2,3} & a_{2,4} \\
  a_{3,1} & a_{3,2} & a_{3,3} & a_{3,4} 
\end{pmatrix}\begin{pmatrix}
  b_{1,1} & b_{1,2} & b_{1,3} & b_{1,4} & b_{1,5} & b_{1,6} \\
  b_{2,1} & b_{2,2} & b_{2,3} & b_{2,4} & b_{2,5} & b_{2,6} \\
  b_{3,1} & b_{3,2} & b_{3,3} & b_{3,4} & b_{3,5} & b_{3,6} \\
  b_{4,1} & b_{4,2} & b_{4,3} & b_{4,4} & b_{4,5} & b_{4,6} 
\end{pmatrix}
\]
can be carried out by computing the matrix product
\[
\begin{pmatrix}
  a_{1,1} & a_{1,2} & a_{1,3} & a_{1,4} \\
  a_{2,1} & a_{2,2} & a_{2,3} & a_{2,4} \\
  a_{3,1} & a_{3,2} & a_{3,3} & a_{3,4} 
\end{pmatrix}\begin{pmatrix}
  b_{1,1} & b_{1,2} & b_{1,3} & b_{1,4} & b_{1,5} \\
  b_{2,1} & b_{2,2} & b_{2,3} & b_{2,4} & b_{2,5} \\
  b_{3,1} & b_{3,2} & b_{3,3} & b_{3,4} & b_{3,5} \\
  b_{4,1} & b_{4,2} & b_{4,3} & b_{4,4} & b_{4,5} 
\end{pmatrix}
\]
and the expressions
\begin{alignat*}1
  &a_{1,1}b_{1,6} + a_{1,2}b_{2,6} + a_{1,3}b_{3,6} + a_{1,4}b_{4,6}\\
  &a_{2,1}b_{1,6} + a_{2,2}b_{2,6} + a_{2,3}b_{3,6} + a_{2,4}b_{4,6}\\
  &a_{3,1}b_{1,6} + a_{3,2}b_{2,6} + a_{3,3}b_{3,6} + a_{3,4}b_{4,6}
\end{alignat*}
which require 12 multiplications.

The additional expressions amount to a matrix multiplication of format
$(3,4,1)$, so the extension can be viewed as obtaining the new $(3,4,6)$ scheme
by patching together a multiplication scheme of format $(3,4,5)$ with one of
format $(3,4,1)$.

Besides the possible alternatives to obtain $(3,4,6)$ from $(3,3,6)$ and
$(3,1,6)$ or from $(2,4,6)$ and $(1,4,6)$, there are further options: we can
also obtain $(3,4,6)$ as a combination of $(3,4,2)$ and $(3,4,4)$, as a
combination of two copies of $(3,4,3)$, or as a combination of two copies of
$(3,2,6)$. 

For all formats $(n,m,p)$ with $n+m+p\leq19$, starting with $(2,2,2)$, we have
performed flip graph searches for starting points that were obtained by
combining two smaller formats in all possible ways. Because of the large number
of combinations, we did not consider increasing the number of start points
further by also using projections, and we also did not keep track of the full
paths that led to improvements. In fact, in this approach it is no longer
adequate to talk about `paths' because every starting point now has two
predecessors rather than one.

One of the formats for which an improvement could be found is $(3,4,6)$. In this
case, the start points obtained from $(3,4,5)$ and $(3,4,1)$ have rank 58 and
could be pushed down to rank 56 via flip graph searches. In contrast, the
starting points obtained from two copies of $(3,2,6)$ have rank 60 but could be
pushed down to rank~54.

In the meta flip graph as defined earlier, if a scheme has several incoming
edges from schemes of other formats, it means that it can be obtained in various
different ways from other schemes by extension or projection. It does not mean
that it was obtained by combining the schemes from which the incoming edges
originate. One way to refine the definition of the meta flip graph such as to
accommodate also the combination of schemes as discussed in the present section
is to introduce a dummy vertex for each such combination. Merging two schemes
$S_1,S_2$ into a new scheme $S_3$ is then expressed edges from $S_1$ and $S_2$
to the dummy vertex and one edge from that vertex to~$S_3$.

\begin{figure}
  \begin{center}
\sffamily
\begin{tikzpicture}[scale=2.5,rounded corners=2pt,inner sep=2pt,outer sep=1pt]
 \node[draw](368) at (0, 0) {\footnotesize368};
 \node(338) at (0, 1) {\footnotesize338};
 \node(337) at (0, 1.5) {\footnotesize337};
 \node(336) at (-1.5, 2) {\footnotesize336};
 \node(237) at (1.5, 2) {\footnotesize237};
 \node(236) at (-2, 3) {\footnotesize236};
 \node(227) at (1.5, 3) {\footnotesize227};
 \node(335) at (-1, 3) {\footnotesize335};
 \node(226) at (-2.5, 4) {\footnotesize226};
 \node(334) at (-1, 4) {\footnotesize334};
 \node(235) at (2, 4) {\footnotesize235};
 \node(234) at (-2, 5) {\footnotesize234};
 \node(333) at (1.5, 5) {\footnotesize333};
 \node(225) at (.25, 5) {\footnotesize225};
 \node(224) at (-1.5, 6) {\footnotesize224};
 \node(233) at (1.5, 6) {\footnotesize233};
 \node(223) at (.5, 6.5) {\footnotesize223};
 \node(222) at (-1.5, 7) {\footnotesize222};
 \begin{scope}[inner sep=0pt,outer sep=0pt]
 \node[above=of 368,circle,draw](368=338+338) {$+$};
 \node[above=of 338,circle,draw,xshift=-3cm](338=233+336) {$+$};
 \node[above=of 338,circle,draw,xshift=3cm](338=333+335) {$+$};
 \node[above=of 335,circle,draw,xshift=2cm,yshift=.5cm](335=233+333) {$+$};
 \node[above=of 235,circle,draw](235=223+233) {$+$};
 \node[above=of 225,circle,draw](225=222+223) {$+$};
 \node[above=of 224,circle,draw](224=222+222) {$+$};
 \node[above=of 234,circle,draw,xshift=1.5cm](234=223+223) {$+$};
 \node[above=of 334,circle,draw,xshift=2cm](334=233+233) {$+$};
 \node[above=of 337,circle,draw,xshift=-1cm](337=233+335) {$+$};
 \node[above=of 337,circle,draw,xshift=1cm](337=333+334) {$+$};
 \node[above=of 237,circle,draw,xshift=-1.5cm](237=233+234) {$+$};
 \node[above=of 237,circle,draw,xshift=1.5cm](237=223+235) {$+$};
 \node[above=of 227,circle,draw,xshift=-1cm](227=222+225) {$+$};
 \node[above=of 227,circle,draw,xshift=1cm](227=223+224) {$+$};
 \node[above=of 226,circle,draw,yshift=1cm](226=222+224) {$+$};
 \node[above=of 226,circle,draw,xshift=1cm](226=223+223) {$+$};
 \node[above=of 236,circle,draw,xshift=1.25cm](236=233+233) {$+$};
 \node[above=of 236,circle,draw,xshift=0cm](236=223+234) {$+$};
 \node[above=of 336,circle,draw,xshift=0cm](336=233+334) {$+$};
 \node[above=of 336,circle,draw,xshift=2cm](336=333+333) {$+$};
 \end{scope}
 \draw[->](368=338+338)edge(368);
 \draw[->](233)edge[bend left=5pt](338=233+336);
 \draw[->](336)edge(338=233+336);
 \draw[->](338=233+336)edge(338);
 \draw[->](337)edge(338);
 \draw[->](333)edge(338=333+335);
 \draw[->](335)edge[bend left=15pt](338=333+335);
 \draw[->](338=333+335)edge(338);
 \draw[->](334)edge(335);
 \draw[->](233)edge(335=233+333);
 \draw[->](333)edge(335=233+333);
 \draw[->](335=233+333)edge(335);
 \draw[->](235)edge[bend right=2pt](335);
 \draw[->](234)edge[bend left=5pt](235);
 \draw[->](223)edge[bend left=5pt](235=223+233);
 \draw[->](233)edge(235=223+233);
 \draw[->](235=223+233)edge(235);
 \draw[->](225)edge(235);
 \draw[->](224)edge(225);
 \draw[->](222)edge(225=222+223);
 \draw[->](223)edge(225=222+223);
 \draw[->](225=222+223)edge(225);
 \draw[->](224=222+222)edge(224);
 \draw[->](223)edge(224);
 \draw[->](222)edge(223);
 \draw[->](233)edge(234);
 \draw[->](224)edge(234);
 \draw[->](234=223+223)edge(234);
 \draw[->](334=233+233)edge(334);
 \draw[->](234)edge(334);
 \draw[->](333)edge(334);
 \draw[->](233)edge(333);
 \draw[->](233)edge(337=233+335);
 \draw[->](335)edge[bend right=15pt](337=233+335);
 \draw[->](337=233+335)edge(337);
 \draw[->](336)edge(337);
 \draw[->](333)edge(337=333+334);
 \draw[->](334)edge(337=333+334);
 \draw[->](337=333+334)edge(337);
 \draw[->](237)edge(337);
 \draw[->](233)edge[bend right=5pt](237=233+234);
 \draw[->](234)edge(237=233+234);
 \draw[->](237=233+234)edge(237);
 \draw[->](236)edge[bend left=5pt](237);
 \draw[->](223)edge(237=223+235);
 \draw[->](235)edge(237=223+235);
 \draw[->](237=223+235)edge(237);
 \draw[->](227)edge(237);
 \draw[->](226)edge(227);
 \draw[->](222)edge[bend left](227=222+225);
 \draw[->](225)edge(227=222+225);
 \draw[->](227=222+225)edge(227);
 \draw[->](223)edge(227=223+224);
 \draw[->](224)edge(227=223+224);
 \draw[->](227=223+224)edge(227);
 \draw[->](222)edge(226=222+224);
 \draw[->](224)edge(226=222+224);
 \draw[->](226=222+224)edge(226);
 \draw[->](225)edge(226);
 \draw[->](226=223+223)edge(226);
 \draw[->](236=233+233)edge(236);
 \draw[->](223)edge(236=223+234);
 \draw[->](234)edge(236=223+234);
 \draw[->](236=223+234)edge(236);
 \draw[->](235)edge(236);
 \draw[->](226)edge(236);
 \draw[->](233)[bend right=5pt]edge(336=233+334);
 \draw[->](334)edge(336=233+334);
 \draw[->](336=233+334)edge(336);
 \draw[->](335)edge(336);
 \draw[->](336=333+333)edge(336);
 \draw[->](236)edge(336);
 \draw[->](223)edge(233);
 \draw[->](338)edge[bend left](368=338+338);
 \draw[->](338)edge[bend right](368=338+338);
 \draw[->](222)edge[bend left=10pt](224=222+222);
 \draw[->](222)edge[bend right=10pt](224=222+222);
 \draw[->](223)edge[bend left=2pt](234=223+223);
 \draw[->](223)edge[bend right=2pt](234=223+223);
 \draw[->](233)edge[bend left=15pt](334=233+233);
 \draw[->](233)edge[bend right=15pt](334=233+233);
 \draw[->](223)edge[bend left=2pt](226=223+223);
 \draw[->](223)edge[bend right=2pt](226=223+223);
 \draw[->](233)edge[bend left=15pt](236=233+233);
 \draw[->](233)edge[bend right=15pt](236=233+233);
 \draw[->](333)edge[bend left=5pt](336=333+333);
 \draw[->](333)edge[bend right=5pt](336=333+333);
\end{tikzpicture}
  \end{center}
  \caption{Genealogy of the improvement found for $(3,6,8)$.}\label{fig:genealogy}
\end{figure}
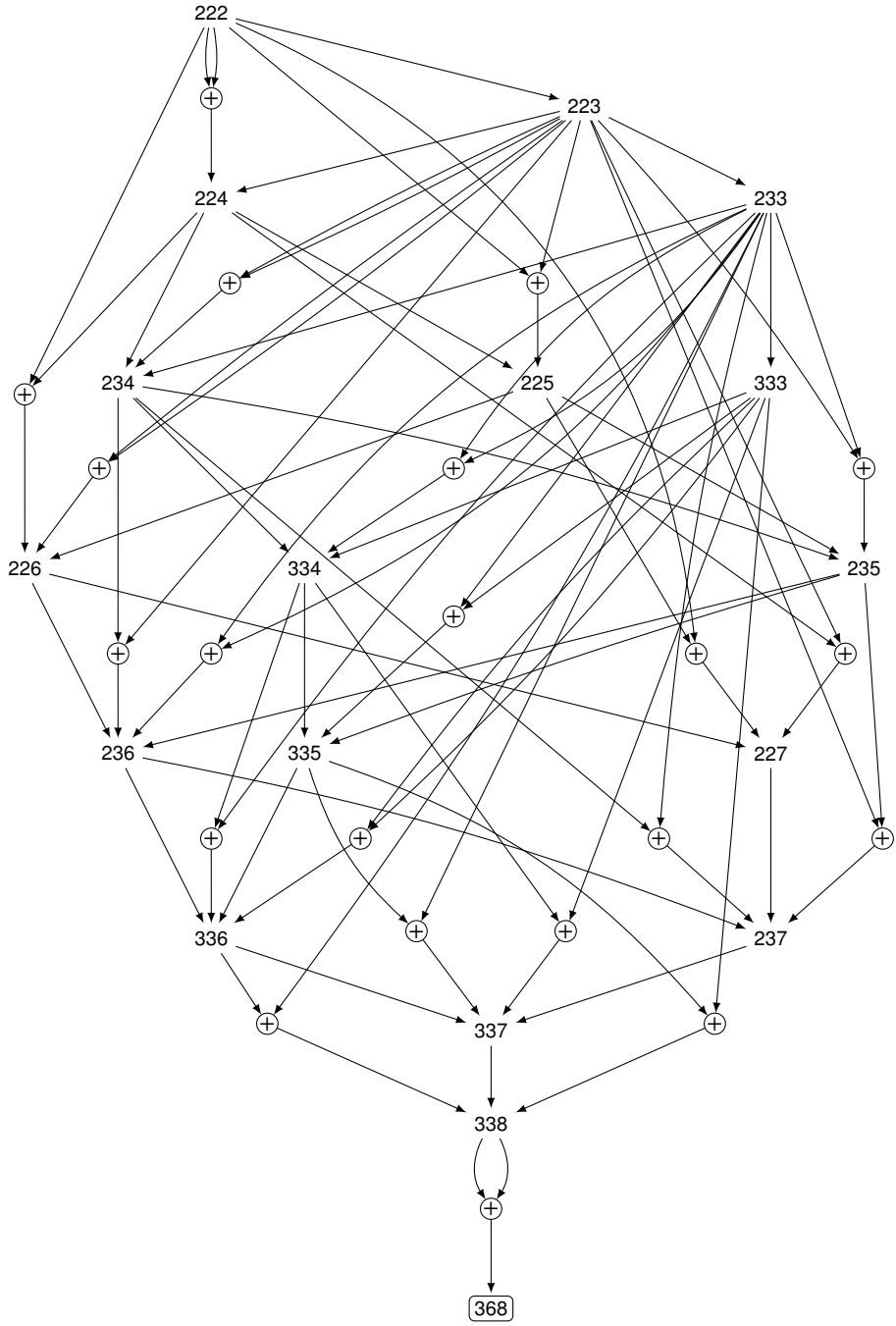

In Figure~\ref{fig:genealogy}, the genealogy of the improvement we found for $(3,6,8)$
is displayed in this sense. The starting points for the successful flip graph search
for this format were obtained by patching together two schemes of format $(3,3,8)$.
Other combinations were examined as well but didn't lead to an improvement and are
therefore not shown in the picture. For the format $(3,3,8)$ itself, the flip graph
search has reached schemes of best rank for starting points obtained in three
different ways: by extensions of schemes of format $(3,3,7)$, by combination
of a scheme of format $(3,3,6)$ with a scheme of format $(3,3,2)$ (taking account
that these schemes are equivalent to schemes of format $(2,3,3)$), and by
combination of a scheme of format $(3,3,5)$ with a scheme of format $(3,3,3)$.
All these best-rank schemes together formed the pool of starting points on which
the search for $(3,3,8)$ was performed. 

\section{Summary of Results}

\begin{figure}
\begin{center}
  \def\o#1{#1}
  \def\b#1{\textbf{#1}}
  \def\c#1{$>$}
  \def\e#1{$=$}
  \def\;{\kern8.3pt}
  \small
  \begin{tabular}{@{}c@{\;}c|c|c@{\;}c@{\;}c@{\;}c@{\;}c@{\;}c@{\;}c@{\;}c@{\;}c|c@{\;}c|c@{\;}c|c@{}}
    size & old & S.~\ref{sec:2} & \multicolumn{9}{|c|}{Starting from AlphaEvolve's schemes} & \multicolumn{2}{|c|}{M/P} & \multicolumn{2}{|c|}{Sect.~\ref{sec:5}} & E \\
    & rank & & 247 & 248 & 256 & 348 & 356 & 357 & 445 & 456 & 447 & 555 & 666 & & & \\\hline 
    \input table
  \end{tabular}
\end{center}
\caption{Summary of results.}\label{tab:1}  
\end{figure}

A summary of our results is given in the table in Figure~\ref{tab:1}. Each row
in this table indicates a matrix multiplication format for which we were able to
improve the previously best known upper bound on the rank as indicated in the
second column. These numbers are taken from \cite{fastmm}, who maintains a table
with the best-known ranks for all matrix multiplication formats $(n,m,p)$
with $\max(n,m,p)\leq32$. 

The lowest ranks we found for a specific rank are highlighted in
bold face.  Other numbers refer to ranks that are better than previously known
but not as good as the improvements via some other path. An ``$=$'' indicates
that the previously best known upper bound (second column) was reproduced. A
``$>$'' indicates that the search got stuck at a rank larger than the rank in the
second column. White space indicates that the format was not covered in the
respective search. Combinatorial explosion limits the number of cases
that can be reached with a reasonable amount of computation time in
approaches that try out all paths. 
  
``S.~\ref{sec:2}'' refers to the search described in Section~\ref{sec:2}.
``Sect.~\ref{sec:5}'' refers to the search described in Section~\ref{sec:5}.
This setting was run twice, thus we give two columns.
The columns labeled ``AlphaEvolve'' and ``M/P'' (for Moosbauer/Poole)
refer to the searches described in Section~\ref{sec:3}.
The column ``E'' refers to results of various experiments we performed in an
early stage of this project. They were partly obtained as described in
Section~\ref{sec:2} and partly obtained using the schemes of Moosbauer and
Poole as starting points, but we are unable to reproduce along which paths
these results were found. We include this column nevertheless because it
contains some improvements that were not found by any of the other searches.

It is noteworthy that each approach finds at least one improvement that each
of the other approaches misses. For the approach of Section~\ref{sec:2}, it
is the improvement for $(2,6,8)$ from 76 to 75. For the approach of Section~\ref{sec:3},
it is for instance the improvement for $(4,5,8)$ from 122 to 118. Interestingly
however, all the improvements we found starting from one of AlphaEvolve's schemes
was also found in some other way. The approach of Section~\ref{sec:5} was most
effective.

As explained in Section~\ref{sec:4}, we were not able to find any improvements
running the flip graph method for some ground field other than $\set Z_2$.  All
the results documented in the table were thus first obtained modulo~2. In most
of the cases, at least one scheme of smallest rank could be lifted to schemes
with coefficients in~$\set Z$, but not always.
For the formats $(4,5,6)$, $(3,7,8)$ and $(3,8,8)$, the improvement appears
to work only over~$\set Z_2$. The other exceptional
formats are $(2,5,7)$, $(2,5,8)$, $(3,4,8)$, $(3,5,7)$, $(2,6,8)$, $(2,7,7)$,
and $(3,7,7)$, where Hensel lifting did not lead to
integer coefficients but only to coefficients in~$\set Q$.

More precisely, for $(2,5,8)$, we found a scheme with coefficients in $\set Z[\frac13,\frac15,\frac17]$,
for $(2,6,8)$, our scheme has coefficients in $\set Z[\frac13,\frac15,\frac1{11}]$,
and for $(3,4,8)$, we even need denominators with twelve distinct prime factors.
For all other formats where we managed to obtain rational coefficients, these
turn out to be in $\set Z[\frac13]$. For $(3,7,7)$, there is in addition one
scheme with coefficients in $\set Z[\frac15]$.

\medskip
The multiplication schemes announced in this article are publicly
available at \url{https://github.com/mkauers/matrix-multiplication}.

\bibliographystyle{plainnat}
\bibliography{bib}

\begin{thebibliography}{19}
\providecommand{\natexlab}[1]{#1}
\providecommand{\url}[1]{\texttt{#1}}
\expandafter\ifx\csname urlstyle\endcsname\relax
  \providecommand{\doi}[1]{doi: #1}\else
  \providecommand{\doi}{doi: \begingroup \urlstyle{rm}\Url}\fi

\bibitem[Arai et~al.(2024)Arai, Ichikawa, and Hukushima]{arai24}
Yamato Arai, Yuma Ichikawa, and Koji Hukushima.
\newblock Adaptive flip graph algorithm for matrix multiplication.
\newblock In \emph{Proc. ISSAC'24}, pages 292--298, 2024.

\bibitem[Drevet et~al.(2011)Drevet, Islam, and
  Schost]{DBLP:journals/tcs/DrevetIS11}
Charles{-}{\'{E}}ric Drevet, Md.~Nazrul Islam, and {\'{E}}ric Schost.
\newblock Optimization techniques for small matrix multiplication.
\newblock \emph{Theor. Comput. Sci.}, 412\penalty0 (22):\penalty0 2219--2236,
  2011.

\bibitem[Dumas et~al.(2025{\natexlab{a}})Dumas, Pernet, and
  Sedoglavic]{dumas25}
Jean-Guillaume Dumas, Cl{\'e}ment Pernet, and Alexandre Sedoglavic.
\newblock A non-commutative algorithm for multiplying a $3\times4$ matrix by a
  $4\times7$ matrix using 63 non-complex multiplications.
\newblock hal-05121550, 2025{\natexlab{a}}.

\bibitem[Dumas et~al.(2025{\natexlab{b}})Dumas, Pernet, and
  Sedoglavic]{dumas25a}
Jean-Guillaume Dumas, Cl{\'e}ment Pernet, and Alexandre Sedoglavic.
\newblock A non-commutative algorithm for multiplying $4\times4$ matrices using
  48 non-complex multiplications.
\newblock Technical Report 2506.13242, ArXiv, 2025{\natexlab{b}}.

\bibitem[Fawzi et~al.(2022)Fawzi, Balog, Huang, Hubert, Romera-Paredes,
  Barekatain, Novikov, Ruiz, Schrittwieser, Swirszcz, Silver, Hassabis, and
  Kohli]{FBH+:Dfmm}
Alhussein Fawzi, Matej Balog, Aja Huang, Thomas Hubert, Bernardino
  Romera-Paredes, Mohammadamin Barekatain, Alexander Novikov, Francisco J.~R.
  Ruiz, Julian Schrittwieser, Grzegorz Swirszcz, David Silver, Demis Hassabis,
  and Pushmeet Kohli.
\newblock Discovering faster matrix multiplication algorithms with
  reinforcement learning.
\newblock \emph{Nature}, 610\penalty0 (7930):\penalty0 47--53, 2022.
\newblock \doi{10.1038/s41586-022-05172-4}.

\bibitem[Heule et~al.(2019)Heule, Kauers, and Seidl]{heule19a}
Marijn~J.H. Heule, Manuel Kauers, and Martina Seidl.
\newblock Local search for fast matrix multiplication.
\newblock In \emph{Proceedings of SAT'19}, volume 11628 of \emph{LCNS}, pages
  155--163, 2019.

\bibitem[Heule et~al.(2021)Heule, Kauers, and Seidl]{HKS:Nwtm}
Marijn~J.H. Heule, Manuel Kauers, and Martina Seidl.
\newblock New ways to multiply $3 \times 3$-matrices.
\newblock \emph{J. Symbolic Comput.}, 104:\penalty0 899--916, 2021.
\newblock ISSN 0747-7171.
\newblock \doi{10.1016/j.jsc.2020.10.003}.

\bibitem[Hopcroft and Kerr(1971)]{HK:OMtN}
J.~E. Hopcroft and L.~R. Kerr.
\newblock On minimizing the number of multiplications necessary for matrix
  multiplication.
\newblock \emph{SIAM Journal on Applied Mathematics}, 20\penalty0 (1):\penalty0
  30--36, 1971.
\newblock \doi{10.1137/0120004}.

\bibitem[Kauers and Moosbauer(2023)]{kauers23f}
Manuel Kauers and Jakob Moosbauer.
\newblock Flip graphs for matrix multiplication.
\newblock In \emph{Proc. ISSAC'23}, pages 381--388, 2023.

\bibitem[Kauers and Moosbauer(2025)]{kauers24c}
Manuel Kauers and Jakob Moosbauer.
\newblock Some new non-commutative matrix multiplication algorithms of size
  $(n,m,6)$.
\newblock \emph{Communications in Computer Algebra}, 58\penalty0 (1):\penalty0
  1--11, 2025.

\bibitem[Kauers and Wood(2025)]{kauers25}
Manuel Kauers and Isaac Wood.
\newblock Consequences of the {M}oosbauer-{P}oole algorithms.
\newblock \emph{Communications in Computer Algebra}, 2025.
\newblock to appear.

\bibitem[Laderman(1976)]{La:Anaf}
Julian~D. Laderman.
\newblock A noncommutative algorithm for multiplying $3\times 3$ matrices using
  23 multiplications.
\newblock \emph{Bull. Amer. Math. Soc.}, 82\penalty0 (1):\penalty0 126--128,
  1976.
\newblock ISSN 0002-9904.
\newblock \doi{10.1090/S0002-9904-1976-13988-2}.

\bibitem[Moosbauer and Poole(2025)]{moosbauer25}
Jakob Moosbauer and Michael Poole.
\newblock Flip graphs with symmetry and new matrix multiplication schemes.
\newblock In \emph{Proc. ISSAC'25}, 2025.
\newblock to appear.

\bibitem[Novikov et~al.(2025)Novikov, Vu, Eisenberger, Dupont, Huang, Wagner,
  Shirobokov, Kozlovskii, Ruiz, Mehrabian, Kumar, See, Chaudhuri, Holland,
  Davies, Nowozin, Kohli, and Balog]{novikov25}
Alexander Novikov, Ngan Vu, Marvin Eisenberger, Emilien Dupont, Po-Sen Huang,
  Adam~Zsolt Wagner, Sergey Shirobokov, Borislav Kozlovskii, Francisco J.~R.
  Ruiz, Abbas Mehrabian, M.~Pawan Kumar, Abigail See, Swarat Chaudhuri, George
  Holland, Alex Davies, Sebastian Nowozin, Pushmeet Kohli, and Matej Balog.
\newblock Alphaevolve: A coding agent for scientific and algorithmic discovery.
\newblock Technical Report 2506.13131, ArXiv, 2025.

\bibitem[Pan(1984)]{pan84}
Victor Pan.
\newblock \emph{How to Multiply Matrices Faster}, volume 179 of \emph{LNCS}.
\newblock Springer, 1984.

\bibitem[Sedoglavic(2017)]{sedoglavic17}
Alexandre Sedoglavic.
\newblock A non-commutative algorithm for multiplying ($7\times7$) matrices
  using 250 multiplications.
\newblock Technical Report 1712.07935, ArXiv, 2017.

\bibitem[Sedoglavic(2025)]{fastmm}
Alexandre Sedoglavic.
\newblock {Yet another catalogue of fast matrix multiplication algorithms}.
\newblock \url{https://fmm.univ-lille.fr/}, 2025.
\newblock Accessed: 2025-04-30.

\bibitem[Smirnov(2013)]{Sm:Tbca}
Alexey~V. Smirnov.
\newblock The bilinear complexity and practical algorithms for matrix
  multiplication.
\newblock \emph{Zh. Vychisl. Mat. Mat. Fiz.}, 53\penalty0 (12):\penalty0
  1970--1984, 2013.
\newblock ISSN 0044-4669.
\newblock \doi{10.1134/S0965542513120129}.

\bibitem[von~zur Gathen and Gerhard(2013)]{vG:MCA}
Joachim von~zur Gathen and J\"{u}rgen Gerhard.
\newblock \emph{Modern Computer Algebra}.
\newblock Cambridge University Press, 3 edition, 2013.
\newblock ISBN 1107039037.

\end{thebibliography}
\end{document}